\documentclass[epj]{webofc}
\usepackage[varg]{txfonts}   
\usepackage{hyperref}
\usepackage{siunitx}
\usepackage{amsmath}

\renewcommand*{\eqref}[1]{Eq.~(\ref{eq:#1})}

\wocname{\includegraphics[width=0.25cm,clip]{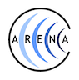} ARENA2018}
\wocname{ARENA2018}
\woctitle{ARENA2018}
\woctitle{\includegraphics[width=0.25cm,clip]{logoARENA} ARENA2018}
\begin{document}
\title{ARIANNA: Measurement of cosmic rays with a radio neutrino detector in Antarctica}
%
%

\author{\firstname{Christian} \lastname{Glaser}\inst{1}\fnsep\thanks{\email{christian.glaser@uci.edu}, CG is supported by the German Research Foundation, (DFG), on grant GL 914/1-1.} for the ARIANNA collaboration}

\institute{Department of Physics and Astronomy, University of California, Irvine, USA}

\abstract{%
  The ARIANNA detector aims to detect neutrinos with energies above \SI{e16}{eV} by instrumenting 0.5 Teratons of ice with a surface array of a thousand independent radio detector stations in Antarctica. The Antarctic ice is transparent to the radio signals caused by the Askaryan effect which allows for a cost-effective instrumentation of large volumes. Several pilot stations are currently operating successfully at the Moore's Bay site (Ross Ice Shelf) and at the South Pole.
As the ARIANNA detector stations are positioned at the surface, the more abundant cosmic-ray air showers are also measured and serve as a direct way to prove the capabilities of the detector. We will present measured cosmic rays and will show how the incoming direction, polarization and electric field of the cosmic-ray pulse can be reconstructed from single detector stations comprising 4 upward and 4 downward facing LPDA antennas. 
}
\maketitle
\section{Introduction}
\label{intro}

The detection of high energy neutrinos requires instrumentation of huge volumes which can be achieved with the radio technique due to the small attenuation of radio signals in ice. 
The ARIANNA neutrino detector exploits this technique and a pilot array of near-surface radio detector stations has been running successfully since several years \cite{HRA, ARENA2018Nelles}. An important physical background is the radio emission of cosmic-ray air showers that is picked up by the in ice antennas. Therefore, a good understanding of cosmic-ray radio signals is crucial for a successful background rejection especially as cosmic-ray signals are several orders of magnitude more abundant. However, cosmic rays are not only a background that we need to get rid of but also a perfect calibration source for a radio-neutrino detector because their radio pulses are very similar to the Askaryan pulses that we expect from neutrinos. Both are bipolar pulses of just a few nanosecond length which are essentially impossible to generate artificially. Therefore, measuring cosmic rays is the only way to fully test a neutrino detector under realistic conditions. Furthermore, the radio emission of air showers is well understood so that the reconstructed signal properties can be verified using theoretical predictions.

\section{Overview of cosmic-ray stations}

\textbf{Dedicated cosmic ray station 32:} This station consists of four upward pointing LPDA antennas that are placed a few meters below the surface into the firn. The excellent timing of the electronics of $\mathcal{O}(\SI{100}{ps})$ allows for a precise reconstruction of the incoming signal direction with an uncertainty of less than \SI{1}{\degree} although the lever arm is only \SI{8}{m}. The two orthogonal polarizations of the antennas allow for a reconstruction of the signal polarization. We note that for neutrinos the polarization is essential to reconstruct the neutrino direction from the signal arrival direction.

\textbf{Neutrino/cosmic-ray hybrid station 52:} This station already comes closer to the optimal design of a radio neutrino detector and combines downward facing LPDAs for neutrino detection with upward pointing LDPAs for efficient cosmic-ray detection/vetoing.

\textbf{Horizontal cosmic-ray station (HCR)} This station consists of eight LPDAs placed above the snow pointing onto the mountain ridge that surrounds Moore's Bay. With this station we study the sensitivity to measure horizontal air showers with the final goal to detect tau neutrinos that interact in the mountains, create taus that escape the mountain and create air showers \cite{HCRICRC17}. 

\textbf{South Pole prototype station 51:} The first ARIANNA station that has been installed at the South Pole. Similar to station 52, it combines downward and upward facing LPDAs complemented by one dipole for a direct measurement of the vertical polarization. The upward facing LPDAs are placed above the surface to better study the RFI noise situation at the South Pole. 

The ARIANNA South Pole station has been operating continuously since its deployment in December 2017, which allowed us to study the (transient) noise condition over an extended period of time. We observe more RFI than at the extremely radio quiet Moore's Bay site. During the summer month in which a lot of human activity takes place at South Pole station, we observed high rates of high amplitude RFI pulses that often saturate our trigger at \SI{10}{Hz}. After South Pole station closed beginning of March and the human activity reduced significantly, we observed both quiet periods of several days followed by noisy periods of several days. A preliminary analysis suggests that most RFI pulses originate from the direction of South Pole station and a nearby wind turbine, and are likely correlated with windspeed, corresponding movement of snow and discharge at large metallic structures. The situation improves significantly when triggering on the downward facing in-ice antennas that are less sensitive to signals coming from the horizon.

\begin{figure}[t]
\centering
\includegraphics[width=0.77\textwidth]{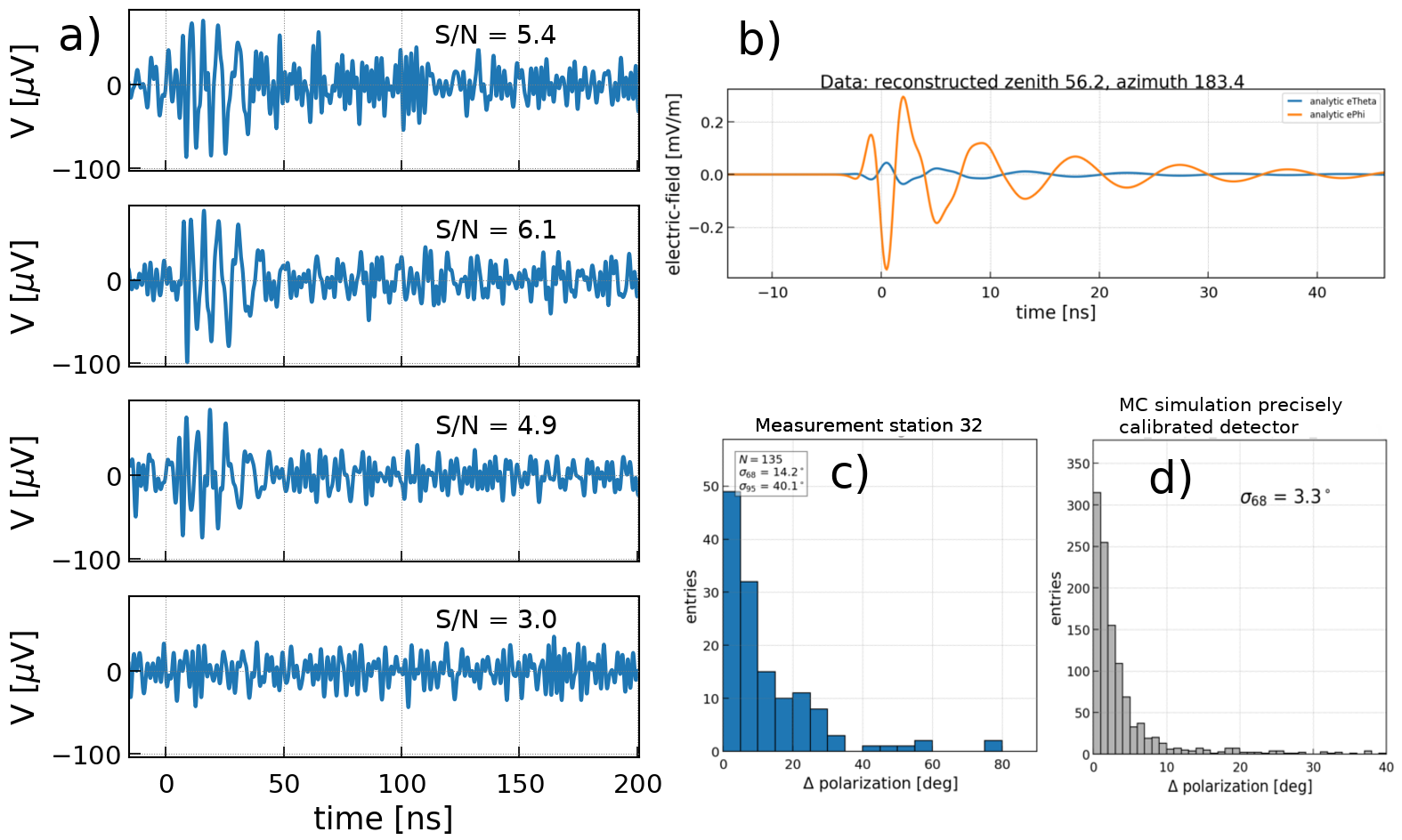}

\caption{a) Air shower measured by the ARIANNA station at South Pole; b) Reconstructed electric field; c) Historgram of difference between measured and expected polarization; d) Expected polarization resolution for a well calibrated detector.}
\label{fig:efield}       
\end{figure}

\section{Polarization reconstruction}
To determine the neutrino direction, both the incoming signal direction and the signal polarization is required because the neutrino radio signal is emitted at an angle of $\sim$\SI{56}{\degree} with respect to the arrival direction on the Cherenkov cone. To break the degeneracy the polarization is needed. We can probe the reconstruction of the signal polarization by measuring air showers. 
We developed a new method to combine the signals from the spatially separated antennas which allows for a model independent unfolding of the antenna response. Hence, we obtain the three-dimensional electric field from which we can calculate the polarization. Furthermore, we developed a new forward folding technique to reconstruct the electric field which improves the reconstruction at small signal-to-noise ratios \cite{GlaserPhD}: We approximate the radio pulse analytically in the frequency domain with just four parameters, the signal amplitude of both polarizations, the frequency slope and a phase offset. We forward fold this pulse with the antenna responses of the different channels and determine the optimal parameters in a chi-square minimization in the time-domain of all channels simultaneously.

We present one example air shower measured at the South Pole station with three LPDAs and one dipole in Fig.~\ref{fig:efield}a. Because the radio signal from cosmic rays is mostly horizontally polarized (the magnetic field in pointing upwards), the dipole does not see any signal. This provides evidence of the cosmic ray origin and provides useful information for the polarization reconstruction. In Fig.~\ref{fig:efield}, we present the reconstructed electric field using the analytic forward folding technique. This technique recovers that the signal is mostly horizontally polarized, which is represented by the orange curve.

We also apply this technique to all data taken by the dedicated cosmic ray station 32 in the 2017/18 season. Cosmic rays are identified using a template matching technique described in \cite{ARIANNACr2017}. For a subset of 135 cosmic-ray events, where the signal amplitude is at least four times the noise RMS in all antennas, we find a good agreement of the reconstructed polarization with the theoretical expectation (cf. Fig.~\ref{fig:efield}c). The resolution is \SI{14}{\degree}, which is expected to improve significantly with an improved detector calibration. To study this, we performed an end-to-end MC simulation using a representative set of CoREAS simulations and including signal distortion due to noise interference. As we do not apply any cut on the distance from the station to the shower core, most events will have signal amplitudes just above the noise threshold, which is a more challenging situation than for typical air shower arrays where air showers are measured by at least 3 stations. For a well-calibrated detector station we find that we can achieve a polarization resolution of 3.3$^\circ$ with the forward folding technique and using both LPDAs and dipoles to measure the signal.

\section{Conclusion}
The detection of cosmic rays with a neutrino radio detector is not only crucial to achieve a good background rejection. It also serves as a tool to calibrate and test the detector under realistic conditions. With the ARIANNA cosmic ray stations we have shown that cosmic rays can be identified reliably using a template matching technique, and that the incoming direction and polarization can be reconstructed. 

%
%
%

\end{document}